\documentclass[aps,pra,twocolumn,superscriptaddress]{revtex4-2}

\usepackage{amsmath,amssymb,graphicx}
\usepackage{color,times}
\usepackage{siunitx}
\usepackage{xcolor}
\definecolor{darkblue}{rgb}{0, 0, 0.8}

\usepackage[colorlinks=true, breaklinks=true, linkcolor=darkblue, citecolor=darkblue, urlcolor=darkblue]{hyperref}

\newcommand{\doilink}[2]{\href{http://dx.doi.org/#1}{#2}}
\newcommand{\reg}[1]{\raisebox{.5pt}{\textcircled{\raisebox{-.9pt} {\footnotesize #1}}}}

\DeclareMathOperator{\erf}{erf}

\begin{document}

\title{\emph{In-situ} equalization of single-atom loading in large-scale optical tweezers arrays}

\author{Kai-Niklas~Schymik} 
\affiliation{Universit\'e Paris-Saclay, Institut d'Optique Graduate School,\\
CNRS, Laboratoire Charles Fabry, 91127 Palaiseau Cedex, France}
\author{Bruno Ximenez}
\affiliation{PASQAL SAS, 7 Rue L{\'e}onard de Vinci, 91300 Massy, France}
\author{Etienne Bloch}
\affiliation{PASQAL SAS, 7 Rue L{\'e}onard de Vinci, 91300 Massy, France}
\author{Davide Dreon}
\affiliation{PASQAL SAS, 7 Rue L{\'e}onard de Vinci, 91300 Massy, France}
\author{Adrien Signoles}
\affiliation{PASQAL SAS, 7 Rue L{\'e}onard de Vinci, 91300 Massy, France}
\author{Florence~Nogrette}
\affiliation{Universit\'e Paris-Saclay, Institut d'Optique Graduate School,\\
CNRS, Laboratoire Charles Fabry, 91127 Palaiseau Cedex, France}
\author{Daniel~Barredo}
\affiliation{Universit\'e Paris-Saclay, Institut d'Optique Graduate School,\\
CNRS, Laboratoire Charles Fabry, 91127 Palaiseau Cedex, France}
\affiliation{Nanomaterials and Nanotechnology Research Center (CINN-CSIC), Universidad de Oviedo (UO), Principado de Asturias, 33940 El Entrego, Spain}
\author{Antoine~Browaeys}
\affiliation{Universit\'e Paris-Saclay, Institut d'Optique Graduate School,\\
CNRS, Laboratoire Charles Fabry, 91127 Palaiseau Cedex, France}
\author{Thierry~Lahaye}
\affiliation{Universit\'e Paris-Saclay, Institut d'Optique Graduate School,\\
CNRS, Laboratoire Charles Fabry, 91127 Palaiseau Cedex, France}

\date{\today}

\begin{abstract}
We report on the realization of large assembled arrays of more than 300 single $^{87}$Rb atoms trapped in optical tweezers in a cryogenic environment at $\sim4$~K. For arrays with $N_{\rm a}=324$ atoms, the assembly process results in  defect-free arrays in $\sim37\%$ of the realizations. To achieve this high assembling efficiency, we equalize the loading probability of the traps within the array using a closed-loop optimization of the power of each optical tweezer, based on the analysis of the fluorescence time-traces of atoms loaded in the traps.
\end{abstract}

\maketitle

Over the last few years, Rydberg atom arrays have emerged as a powerful platform for quantum technologies, with applications ranging from quantum sensing to quantum computing and simulation~\cite{Saffman2016, Madjarov2019, Norcia2019,Browaeys2020,Henriet2020,Morgado2021,Bluvstein2022,Graham2022}. The loading of single atoms in optical tweezers is stochastic, with a probability $\eta$ for a trap to be loaded at any given time. For the usual loading of single alkali atoms from a magneto-optical trap, relying on collisional blockade~\cite{Schlosser2001}, one has $\eta\sim0.5$. Recently, by using more elaborate cooling schemes or alkaline-earth species, higher loading probabilities, ranging form $0.8$ to $0.96$, have been demonstrated~\cite{Grunzweig2010,Sompet2013,Fung2015, Lester2015,Brown2019,Aliyu2021,Jenkins2022}. However, so far, the realization of large, defect-free arrays has to rely on an atom-by-atom assembly procedure, initially demonstrated for a few tens of atoms~\cite{Endres2016,Barredo2016,Kim2016}, and  recently extended to the range of 200 to 300 atoms~\cite{Scholl2021,Ebadi2021,Ebadi2022}. 

\begin{figure}[b]
\begin{center}
\includegraphics[width=85mm]{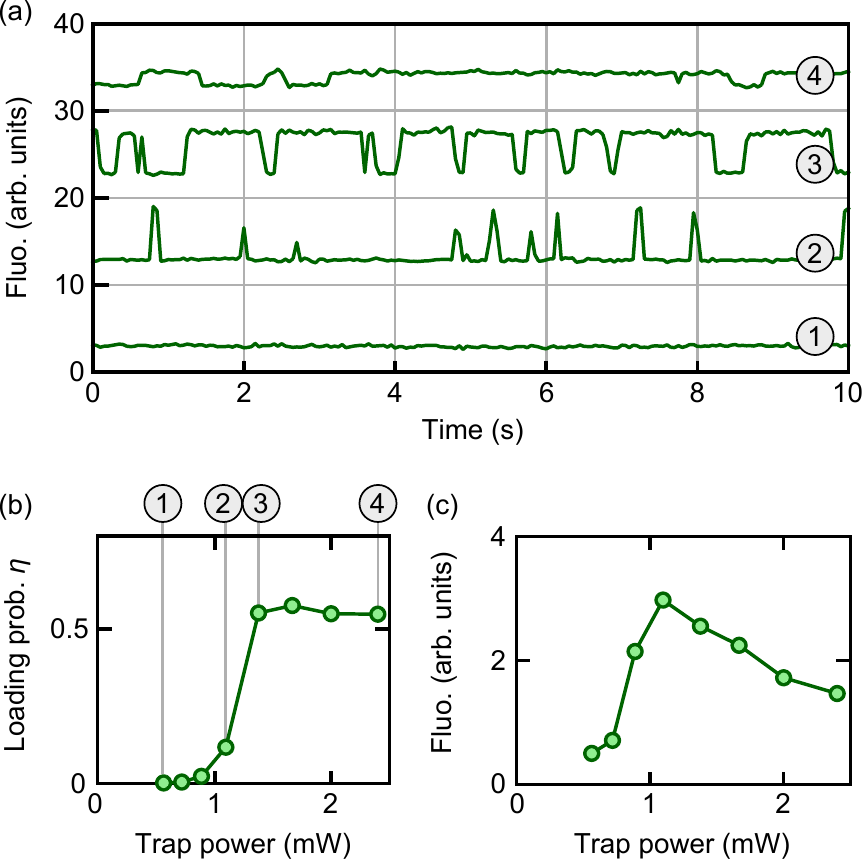}
\caption{Effect of the trap depth on single-atom loading. (a) Time-dependent fluorescence  traces of a single atom for increasing trap powers. The curves have been shifted vertically for clarity. (b) Evolution of the loading probability $\eta$ of a trap as a function of the trap power. (c) Evolution of the amplitude of the fluorescence signal (``fluorescence step'' in the following) of a single atom as a function of the trap power. Ideally, all traps should operate in regime {\reg{3}} to optimize the loading of large arrays.}
\label{fig:fig1}
\end{center}
\end{figure}

Many applications call for scaling the assembly technique to even higher numbers, and significant steps have been made in this direction, with more efficient sorting algorithms~\cite{Schymik2020}, and much longer trapping lifetimes using a  cryogenic setup~\cite{Schymik2021}. However, at the scale of several hundreds of trapped atoms, it becomes increasingly difficult for all optical tweezers in an array to load single atoms efficiently, mostly because the range in trap power over which efficient loading is observed in a trap is quite small (see Fig.~\ref{fig:fig1}). As holographic trap arrays realized with spatial light modulators (SLMs) naturally show a dispersion in the trap intensities, it is necessary to equalize the trap depths in order to achieve an efficient loading throughout the whole array. In large arrays, the difficulty in achieving this is further enhanced by the optical aberrations that appear at the periphery of the field of view of the focusing optics.

In previous work, different methods for trap depth equalization have been used. A first approach, used e.g. in~\cite{Nogrette2014}, consists in imaging the trap array on a diagnostic CCD camera, thus inferring the intensity of each trap, and using this information as the starting point for the calculation of a new hologram with refined target intensities for the traps. However, in large arrays, we observe that this method does not give a sufficient trap-loading performance. Other approaches replace the measurement of the trap intensity using a diagnostic camera by an \emph{in situ} measurement of the light shifts experienced by a trapped atom~\cite{Endres2016,Singh2021,Jenkins2022}. Nevertheless, such an approach can only be applied when the atom loading over the array is already relatively efficient.

In this work, we report on a simple \emph{in-situ} trap-loading equalization technique, based on the analysis of the evolution of single-atom fluorescence traces of all traps of the array as a function of the overall trap power $P_{\rm tot}$. In our setup, it improved drastically the assembly efficiency for large arrays, and allowed us to assemble arrays of more than 300 atoms with an unprecedented probability of $\sim 37\%$ to get defect-free arrays. This article is organized as follows. We first briefly recall the  essential characteristics of our experimental setup, and then describe the procedure used for equalizing the loading probability of optical tweezers in large arrays. We finally demonstrate efficient assembly of arrays up to an unprecedented size of more than 300 atoms, mainly limited by the small field of view ($\pm25\;\mu$m) of our aspheric lenses.

\begin{figure}[t]
\begin{center}
\includegraphics[width=85mm]{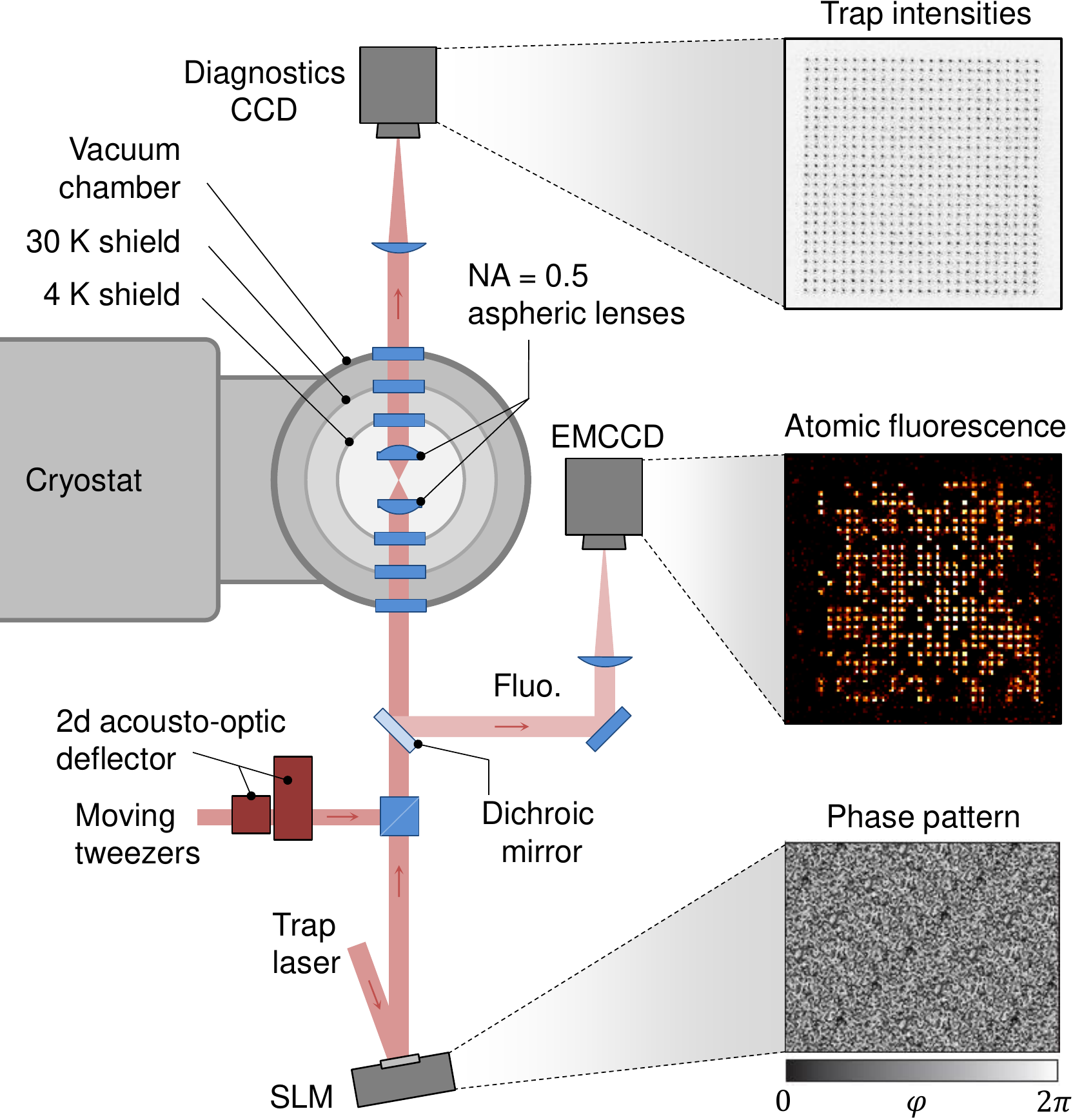}
\caption{Sketch of the experimental setup. The insets show, from bottom to top, the phase pattern used to produce a $23\times23$ square array with a spacing of 5~$\mu$m, an atomic fluorescence image of atoms loaded in the array, and an image of the trap intensities obtained on the diagnostics CCD camera. }
\label{fig:setup}
\end{center}
\end{figure}

Our cryogenic experimental setup has been described in detail in~\cite{Schymik2021} and is sketched in Fig.~\ref{fig:setup}. In brief, we use a closed-cycle, UHV compatible cryostat at 4~K to achieve extremely high vacuum levels in our apparatus, reaching $\sim6000$~s lifetimes for $^{87}$Rb atoms trapped in optical tweezers. The science chamber contains two aspheric lenses (numerical aperture ${\rm NA}=0.5$, focal length 10~mm, working distance 7~mm) allowing us to focus down the trapping beam to a $1/e^2$ radius of about $1\,\mu$m. 

The trapping light is generated by a Titanium-Sapphire laser operating at 815~nm. A single mode fiber is used for light delivery on the setup, and, after diffraction on the SLM, up to $P_{\rm tot}=1.8$~W can be sent into the cryostat (whose base temperature then increases from 4.2~K to 5.8 K). For the SLM hologram calculation, we use the Weighted Gerchberg-Saxton (WGS) algorithm~\cite{DiLeonardo2007,Kim2019} and create arrays containing $N_{\rm t}$ traps, whose individual intensity is controlled by adjustable weights $w_i$ (initially taken all equal).

The quality of the trap arrays can be assessed using a diagnostics CCD camera onto which the array is imaged using the second aspheric lens. We note however that some of the optical aberrations introduced by the focusing lens, such as coma, can be compensated for by the second one and thus do not appear on the diagnostics CCD image. Still, in a first step towards obtaining homogeneous loading throughout the array, we run an ``intensity equalization'' iterative procedure~\cite{Nogrette2014} using the intensities measured on the diagnostics CCD camera to calculate new holograms that improve the homogeneity of the array. For the arrays used here, the relative standard deviation of the trap intensities as measured on the diagnostics CCD camera is $<5\%$ after 2 iterations, and $\sim2\%$ after 5 iterations~\cite{note}. 

The atomic fluorescence is collected through the first asphere (used to focus down the tweezers), separated from the trap light with a dichroic mirror, and imaged onto an electron-multiplication (EMCCD) camera. For a single tweezer, the quality of single-atom trapping depends strongly on the power of the trap, as can be seen in Fig~\ref{fig:fig1}(a). When the trap power is too small (regime {\reg{1}}), one cannot trap any atom. Then, for a slightly higher power, some single atoms are occasionally trapped, but they spend little time in the tweezers, yielding a very low occupancy $\eta$ of the trap (regime {\reg{2}}). For a higher power in the tweezers (typically around 1.5~mW for our trap parameters, corresponding to a trap depth of about $k_B\times1$~mK) one achieves at the same time a high loading probability $\eta\simeq55\%$, and a high value of the fluorescence signal that makes it easy to discriminate between the presence and the absence of an atom (regime {\reg{3}}). For even higher powers, while the loading efficiency $\eta$ remains roughly constant, the increased light-shift experienced by the atoms in the tweezers reduces the fluorescence step size (regime {\reg{4}}). Finally, at very large powers (not shown) the tweezers start to accommodate several atoms as light-assisted collisions become inefficient. We ideally want to work in regime {\reg{3}}, which corresponds to the minimal power for reaching a high $\eta$, thus allowing the realization of the largest number of traps $N_{\rm t}$  for a given trap laser power $P_{\rm tot}$.

\begin{figure}[t]
\begin{center}
\includegraphics[width=80mm]{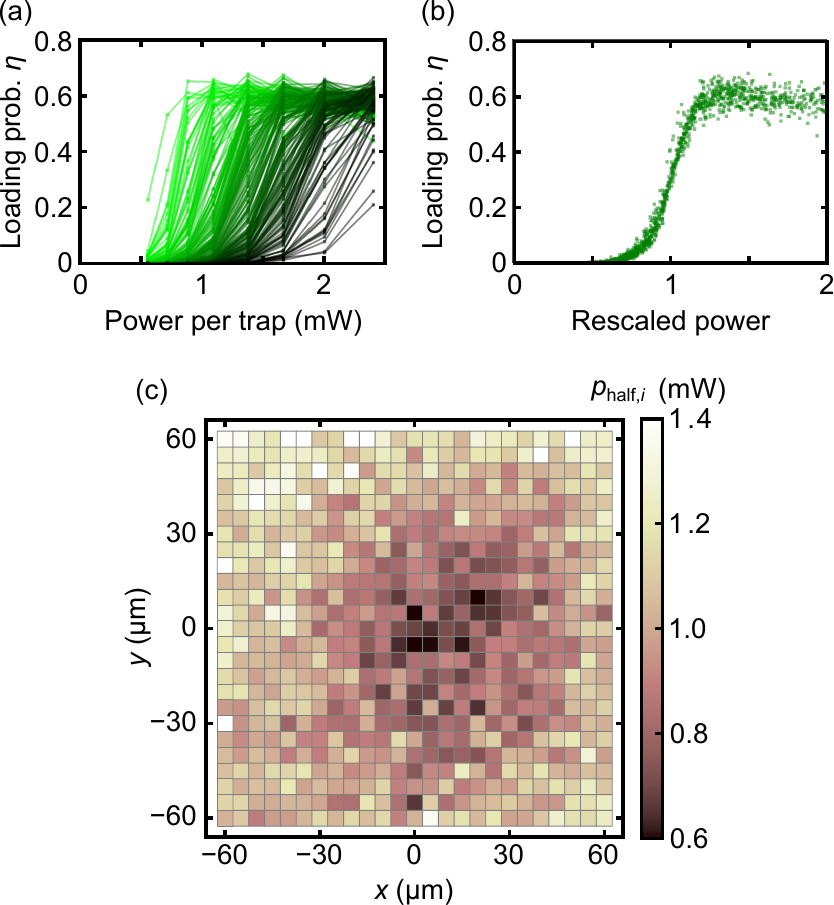}
\caption{Inhomogeneity of the trap intensities in large square array. (a):   Loading efficiency $\eta$ as a function of the average power per trap $P_{\rm tot}/N_{\rm t}$, showing a large dispersion in a $15\times15$ array. (b): The same data collapse on a universal curve when the $x$-axis is rescaled by $p_{{\rm half},i}$ obtained from a fit by Eqn. (\ref{eq:fit}) of all individual curves of panel (a). (c): Spatial distribution of the fitted values of $p_{{\rm half},i}$ for the case of a $25\times25$ array, showing that, due to optical aberrations, traps in the periphery require more power to efficiently load single atoms.}
\label{fig:fig3}
\end{center}
\end{figure}

Despite having performed the intensity equalization procedure described above, we observe on large arrays that even for an empirically optimized value of $P_{\rm tot}$, some traps are  still in regime {\reg{1}} while others may already be in regime {\reg{4}}. To study quantitatively this variation of the loading efficiency, we fix $P_{\rm tot}$ and record,  for a duration of typically $30$~s in order to have enough statistics, the fluorescence time traces of all traps. We then extract for each trap $i$ its loading probability $\eta_i$ (measured as the fraction of time when an atom resides in the trap). Figure~\ref{fig:fig3}(a) shows the resulting curves $\eta_i(P_{\rm tot}/N_{\rm t})$, which all exhibit the same global shape seen in Fig.~\ref{fig:fig1}(b), but with an overall scaling along the $x$-axis. We make this statement more quantitative by fitting each individual curve by an error-function
\begin{equation}
\eta_i = \frac{\eta_i^{\rm max}}{2} \left\{\erf\left[\alpha\left(\frac{P_{\rm tot}}{N_{\rm t}}-p_{{\rm half},i}\right)\right]+1\right\},
\label{eq:fit}
\end{equation}
where the parameter $p_{{\rm half},i}$ allows us to locate the average power per trap at which trap $i$ reaches half its asymptotic value $\eta_i^{\rm max}$; the parameter $\alpha$ accounts for the width of the transition region. Then, when plotting the loading curves as a function of the power per trap \emph{rescaled} by $p_{{\rm half},i}$, we observe a collapse of all the data points on a universal curve, as can be seen in Fig.~\ref{fig:fig3}(b). Achieving a power of about $1.5p_{\rm half}$ on each trap would result in an optimal loading. A similar universal behavior is observed, after rescaling, for the size of the fluorescence steps.

It is instructive to see how the value of $p_{{\rm half},i}$ correlates with the position of trap $i$. Figure~\ref{fig:fig3}(c) shows the extracted values of $p_{{\rm half},i}$ as a function of the trap positions in a $25\times25$ array, revealing that traps in the periphery of the optical system require, on average, twice as much power as those in the center in order to trap atoms optimally. We attribute this effect to optical aberrations, the size of the array being much bigger than the specified  field of view ($\pm25\;\mu$m) of the aspheric lenses. In addition to these large-scale variations, we also observe significant fluctuations from to trap to trap.

\begin{figure}[t]
\begin{center}
\includegraphics[width=85mm]{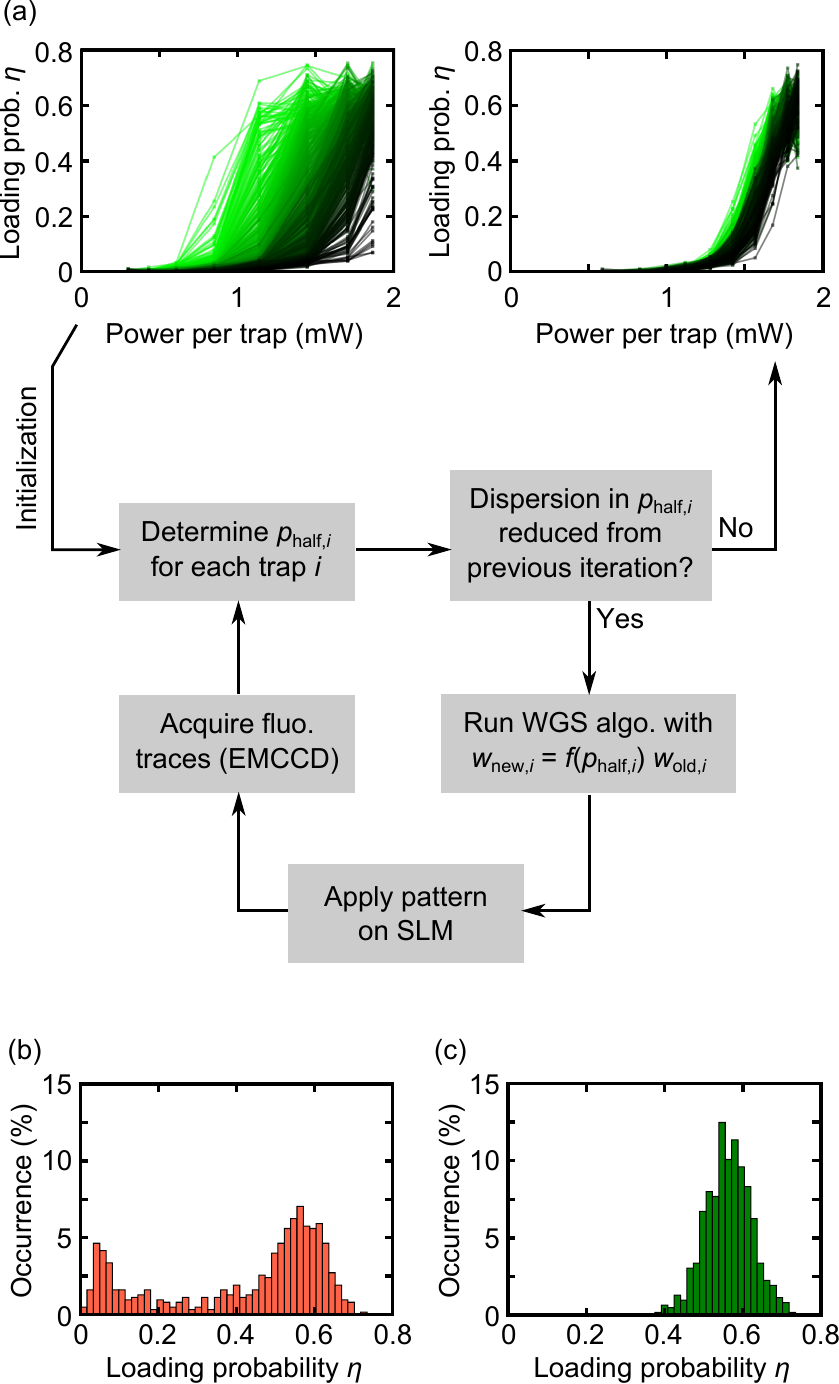}
\caption{(a): The loading equalization procedure on a $25\times25$ array (see text for the definition of the function $f$). Starting from a broad distribution of loading curves $\eta_i$ (top left; the standard deviation of the fitted $p_{{\rm half},i}$ is 17\% over the array), we obtain, after 5 iterations of the loop, a much narrower distribution (top right; the standard deviation of the fitted $p_{{\rm half},i}$ is 3\%). Distribution of the loading probabilities $\eta_i$ over the array before (b) and after (c) the equalization procedure, for a total power $P_{\rm tot}=N_{\rm t}\times 1.7$~mW.} 
\label{fig:fig4}
\end{center}
\end{figure}

\begin{figure}[t]
\begin{center}
\includegraphics[width=80mm]{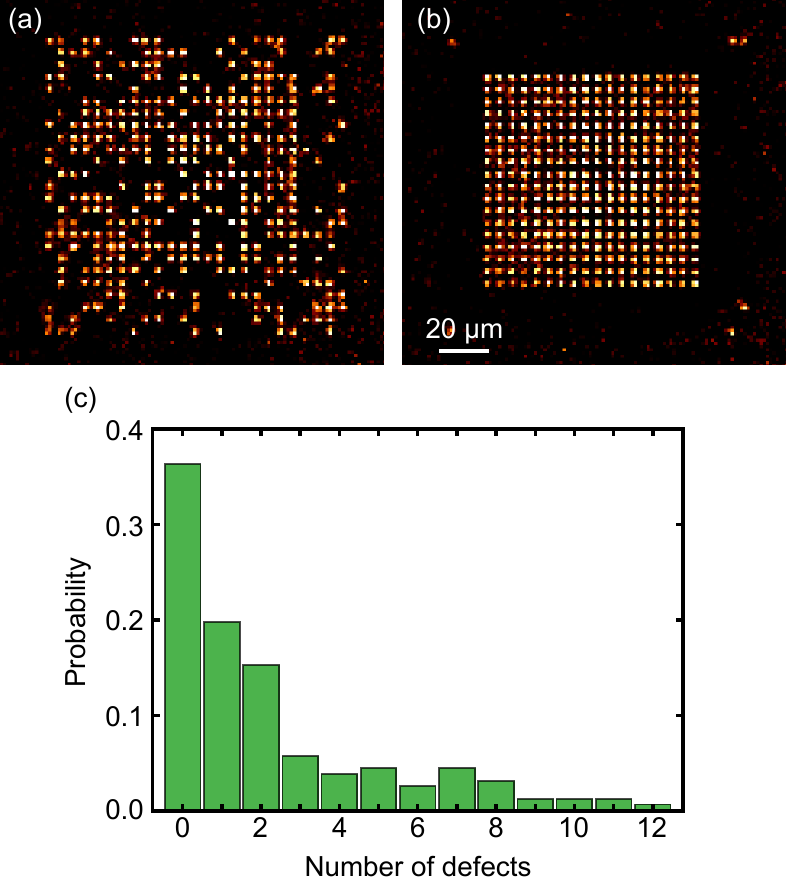}
\caption{Efficient assembly of a 324-atom array. (a): Fluorescence image of the 625-trap array before rearrangement. The lattice spacing is 5~$\mu$m. (b): Fluorescence image of the array after two rearrangement cycles, showing a defect-free 324-atom array together with remaining reservoir atoms that have not been dumped yet. (c): Probability distribution of the number of missing atoms in the target array, showing a large ($\simeq37$\%) probability of preparing \emph{defect-free} arrays.}
\label{fig:fig5}
\end{center}
\end{figure}

We  now harness this \emph{in-situ} characterization of the loading efficiency of each trap to calculate improved holograms, that distribute the light unevenly among traps and  thus directly optimize the uniformity of the trap loading over the entire array.

To do so, we follow the procedure outlined in Fig.~\ref{fig:fig4}. We analyze the evolution with $P_{\rm tot}$ of the loading probability $\eta_i$ of each trap, and extract the values $p_{{\rm half},i}$: a large value of this parameter for a given trap means that in the next SLM pattern calculation, one should target a higher relative intensity for that trap. This analysis takes a few minutes. We then calculate a new hologram with updated intensity weights 
\begin{equation}
w_{{\rm new},i}=f(p_{{\rm half},i})w_{{\rm old},i}.
\end{equation} 
We normally chose the function $f$ to be proportional to $p_{{\rm half},i}$ while keeping a constant total power, i.e. 
\begin{equation}
f(p_{{\rm half},i}) = N_{\rm t} p_{{\rm half},i}/P_{\rm tot}.
\label{eq:simple:f}
\end{equation}
However, for our largest arrays, we found that using this simple functional form can  lead to an oscillatory behavior in the optimization loop, and thus we use a modified function
\begin{equation}
f(p_{{\rm half},i}) = \frac{1}{1-G\left[1-P_{\rm tot}/(N_{\rm t} p_{{\rm half},i})\right]},
\end{equation}
where we adjust the ``gain'' $G$ slightly below 1 to optimize convergence (for $G=1$ we recover Eqn.~(\ref{eq:simple:f})).
 
Typically ten iterations of the WGS algorithm are then run, which takes another few minutes, after which the trap intensities are measured to be within $\sim5\%$ of the requested value. Then a new cycle [fluorescence analysis$/$hologram calculation] can take place. We find that when we start from an entirely new array of traps, 5 to 8 such cycles are needed to converge to a situation where the distribution of $p_{{\rm half},i}$ has become quite narrow and does not evolve any more, meaning that the optimization procedure can be completed in about one hour. Figure \ref{fig:fig4} (b) and (c) show the distributions of loading probabilities $\eta_i$ in the array, before and after running the optimization loop, respectively. The improvement in the loading uniformity is drastic, with all traps showing a loading probability $\eta>0.4$. We have observed that once performed, the loading remains optimal over many days in our laboratory environment.

We finally illustrate the efficiency of the equalization procedure by studying the rearrangement of large arrays. Figure~\ref{fig:fig5}(a,b) shows the realization of a fully-loaded array with $N_{\rm a}=324$~atoms within an array of $N_{\rm t}=625$ traps. We use the `LSAP-2' atom-sorting algorithm that we developed in~\cite{Schymik2020}; the typical number of individual atom moves is $\sim300$, each lasting about $1$~ms. Figure~\ref{fig:fig5}(c) shows the probability distribution of the number of defects in the target array. In about 37\% of the shots, a defect-free array is obtained. For comparison, in our room-temperature setup and without the \emph{in-situ} equalization method discussed here, we could only achieve in~\cite{Scholl2021} a 3\% probability of defect-free $N_{\rm a}=196$ arrays. Pushing our experimental setup to its limits in terms of laser power, we have been able to assemble arrays with up to $N_{\rm a}=361$~atoms (not shown). We believe that this number can be significantly increased by using an optical system with a larger field of view, such as a microscope objective.

In conclusion, we have demonstrated a simple procedure allowing us to optimize the loading of holographic trap arrays, using only a simple analysis of the fluorescence time-traces of single atoms loaded in the tweezers array.  Ultimately, the overall assembly efficiency is limited by two factors. The first one are the losses during the transfer of an atom with the moving tweezers from a source to a target trap; currently the loss probability, averaged over all possible moves, is $\simeq1\%$.  The second limitation arises from losses that occur during the imaging, and are currently at the level of $\simeq0.2\%$, again averaged over the entire array. A detailed study of both limitations will be the subject of future work. 

\begin{acknowledgments}
We thank P. Scholl and S. Pancaldi for contributions in the early stages of this study. KNS acknowledges funding from the Studienstiftung des deutschen Volkes, and DB from MCIN/AEI/10.13039/501100011033 (RYC2018-025348-I, PID2020-119667GA-I00, and European Union NextGenerationEU PRTR-C17.I1). This project has received funding from R\'egion \^Ile-de-France through the DIM SIRTEQ (project CARAQUES), from the European Union's Horizon 2020 research and innovation program under grant agreement no. 817482 (PASQuanS), from the ERC Advanced grant no. 101018511 (ATARAXIA), and from Bpifrance (Proqure project).
\end{acknowledgments}


\begin{thebibliography}{99}
\bibitem{note}
In a setup where the array size is smaller than the field of view of the focusing system, for instance when using microscope objectives, such an intensity equalization can be enough to achieve efficient loading over the array; however, as we shall see below, due to the modest field of view of our aspheric lenses, arrays with several hundreds of traps do not achieve a homogeneous loading in our setup with intensity equalization alone.
\end{thebibliography}
\end{document}